# Robust single-sideband-modulated Raman light generation for atom interferometry by FBG-based optical rectangular filtration

Guochao Wang[1,2,4,#,*], Yaning Wang[1,2,#], Kang Ying[3], Huankai Zhang[1,2], Xu Zhang[1,2], Qixue Li[1,2], Xuan Li[3], Enlong Wang[1,2], Xiao Yu[1,2], Aiai Jia[1,2], Shuhua Yan[1,2], Jun Yang[1,2], Lingxiao Zhu[1,2,*]

[1] *College of Intelligence Science and Technology, National University of Defense Technology, Changsha, Hunan 410073, China*

[2] *Interdisciplinary Center for Quantum Information, National University of Defense Technology, Changsha, Hunan 410073, China*

[3] *Shanghai Institute of Optics and Fine Mechanics (SIOM), Chinese Academy of Sciences, Shanghai, 201800, China, yingk0917@siom.ac.cn*

[4] *High-tec Institute of Xi'an, Xi'an 710025, China*

[#] These authors contributed equally to this work.

[*] E-mail address: wgc.19850414@163.com, zhulingxiao31@163.com

**Abstract:** Low-phase-noise and pure-spectrum Raman light is vital for high-precision atom interferometry by two-photon Raman transition. A preferred and prevalent solution for Raman light generation is electro-optic phase modulation. However, phase modulation inherently brings in double sidebands, resulting in residual sideband effects of multiple laser pairs beside Raman light in atom interferometry. Based on a well-designed rectangular fiber Bragg grating and a plain electro-optic modulator, optical single-sideband modulation has been realized at 1560 nm with a stable suppression ratio better than -25 dB despite of intense temperature variations. After optical filtration and frequency doubling, a robust phase-coherent Raman light at 780 nm is generated with a stable SNR of better than -19 dB and facilitates measuring the local gravity successfully. This proposed all-fiber single-sideband-modulated Raman light source, characterized as robust, compact and low-priced, is practical and potential for field applications of portable atom interferometry.

## 1. Introduction

Atom interferometry, characterized as a powerful art of coherently manipulating the motion of atoms and surveying atom waves, potentially has ultra-high precision and sensitivity comparing with classical optical interferometry [1]. Atom interferometers play significant roles in atomic physics, quantum information and precision measurement, and have become efficient and mature tools for studying fundamental quantum mechanical phenomena [2-5]. In a stimulated Raman-transition atom interferometer, two counter-propagating Raman light beams, each of which has dual frequencies with a gap tuned to Raman resonance corresponding to a certain radio frequency, are used to stimulate Raman transitions between two ground states. The performance of Raman light directly determines the accuracy and sensitivity of the atom interferometer in terms of signal-to-noise ratio and phase shift for fringe pattern [6]. Hence, Raman light generation with low phase noise and pure spectrum is essential for high precision atom interferometry.

At present, there are three main methods for Raman light generation, namely optical synthesis by phase-locking, optical synthesis by frequency shift, and frequency generation by electro-optic modulation. The phase-locking method is a classical and universal mean, but an intricate phase-locking system with optics and circuits is indispensable to achieve well locking, which has strict requirements for phase noise and bandwidth in the whole loop [7]. The frequency-shifting method is seldom used due to limited frequency shifts and low efficiency of the high-frequency modulation [8]. Compared with the other two methods, frequency generation by electro-optic modulation uses an electro-optic modulator (EOM) to directly generate coherent Raman light with the carrier and sidebands mixed [9][10]. As the modulated Raman light has uniform polarization, equal frequency interval, and tunable power ratio calibrated by the modulation depth, so it is intrinsically coherent with a phase noise comparable with the RF source. These advantages make this method most engineering practical and potential to construct a compact and integrated Raman light source for the portable atom interferometers. However, the modulation method equally has a prominent disadvantage of double sidebands. Consequently, there are residual sidebands apart from the demanded Raman light, resulting in deteriorated fringe contrast of atomic interference and phase shift, which is referred to as the sideband effect [11].

Lots of efforts have been made to eliminate the sideband effect of the electro-optic modulation by groups of atom interferometry worldwide. D. M. S. Johnson and M. A. Kasevich proposed Serrodyne frequency shifting of light from 200 MHz to 1.2 GHz to promote the efficiency of the concerned sideband, however, residual sidebands could be reduced but unprevented [12]. K. S Lee reported a single high-power laser beam composed of two phase-coherent sidebands without the perturbing carrier mode using the Fabry-Perot cavity with a free spectral range of 6.8 GHz [13], but their scheme required sophisticated controls of high-finesse Fabry-Perot cavity and the power ratio of the two laser frequencies of Raman light was fixed to 1:1 unchangeably due to the same intensity of the ±1st order sideband. N. Arias firstly tried to use a high-power phase modulator and a long calcite crystal to rotate the polarization of the sidebands with respect to the carrier. Through elaborate design of the optical path, it converted the destructive interference of the Raman pairs into constructive interference and filtered out the sideband by polarization distinction [14]. L. Zhu initially demonstrated a single-sideband-modulated Raman light based on in-phase and quadrature (IQ) modulation, achieving a remarkable suppression ratio of -21 dB on additional sidebands [15], but additional electronic controls were relatively complex and indispensable. Moreover, resulting from weak immunity against environmental perturbations and drifts of the modulator's working points, its long-term stability needed to be further improved [16]. Recently, C. D. Macrae demonstrated a single-seed, frequency-doubled fiber laser system relied on fiber Bragg grating (FBG) technology for quantum sensing. They used composite RF signals of 14.8 GHz and 7.965 GHz to drive a high-bandwidth EOM claimed as 30 GHz, and filtered out the desired -1st sidebands by a reflective FBG for frequency doubling. As the used FBG failed to suppress one of -2nd sidebands within its bandwidth, multiple undesired frequencies at 780 nm emerged besides Raman light, and an ordinary signal to noise ratio (SNR) of -15 dB was demonstrated [17].

Optical single sideband (OSSB) modulation as one of the most promising photonic techniques can be achieved by FBG filtering [18], which has been widely applied in the fiber optical communication systems, long-distance radio-over-fiber (ROF) system, and microwave photonic systems [19][20]. In this work, relying on the FBG-based OSSB modulation, we propose a practical and efficient method for robust OSSB-modulated Raman light generation with all fiber-communication optics. Based on a well-designed and homemade rectangular FBG, we succeed in obtaining OSSB modulation with a stable suppression ratio better than -25 dB despite of intense temperature variations. Making use of waveguided frequency doubling, we have achieved a low-phase-noise coherent Raman light with SNR better than -19 dB, which is insensitive to temperature disturbance and achieved at a relatively low price compared to other OSSB-modulated Raman light. This method facilitates a robust and practical all-fiber Raman light source potential for field applications of atom interferometry.

**2. Basic principle of rectangular FBG**

When light waves pass through a uniformly periodic FBG with spatially periodic phase distribution formed in the fiber core, light waves that match the Braggs law is about to reflect, where its essential function is concluded as a reflective mirror with a certain optical bandwidth. Generally, there are four critical parameters about this optical bandpass filter according to the theory of FBG [21], namely center wavelength, reflective bandwidth, reflectivity, and side-mode suppression ratio (SMSR), respectively. The center wavelength $\lambda$ for the reflective spectrum is determined by the grating pitch, conforming to the equation of $\lambda = 2n\Lambda$, where $\Lambda$ is the grating pitch and $n$ the refractive index. The reflective bandwidth of FBG characterized as full width at half-maximum (FWHM) can be expressed as

$$\Delta\lambda = \lambda\sqrt{\left(\frac{\Delta n}{2n}\right)^2 + \left(\frac{\Lambda}{L}\right)^2} = 2n\Lambda\sqrt{\left(\frac{\Delta n}{2n}\right)^2 + \left(\frac{\Lambda}{L}\right)^2} \qquad (1)$$

where $\Delta n$ is the induced index modulation depth, and $L$ is the grating length. As $\Delta n/n$ is usually definite, the reflective bandwidth of FBG is dependent on and basically in inverse proportion to $L$, the required bandwidth corresponding to a specific $L$. The reflectivity $R$ for the uniformly periodic FBG is denoted as

$$R = \tanh^2\left(\frac{\pi}{2} \cdot \frac{L}{\Lambda} \cdot \frac{\Delta n}{n}\right) \qquad (2)$$

As the index modulation depth $\Delta n$ of fiber grating is at the level of $10^{-3}$, a high reflectivity above 99% is readily achieved. Thanks to the ultra-high reflectivity, a fairly flat bandpass spectrum within the reflective bandwidth is achieved better than 1dB, resulting from the FBG reflectivity saturation effect. Furthermore, as the side modes generally exist coincided with the FBG main lobe, the FBG apodized technology is used to reduce the side mode oscillation effect of FBG. Finally, based on the mechanism presented above, ideal optical filtering response for single sideband modulation can be expected by a rectangular FBG compatible with a high suppression ratio, a flat spectrum and a narrow bandwidth.

## 3. Experimental setup

### 3.1 Schematic layout

Figure 1 gives the configuration diagram of the single-sideband-modulated Raman light generation. A frequency-stabilized fiber laser with a linewidth of kHz at 1560 nm, whose frequency has been locked at specific position under the consideration of Raman detuning, passes through an EOM (iXblue, MPZ-LN-10), and generates modulated double sidebands in the presence of a microwave power signal with the frequency gap at about 6.834 GHz for Rb atom. The modulated lasers circumferentially enter the well-designed and homemade FBG by a fiber circulator.  Then two components within the reflective bandwidth, the carrier and the +1$^{st}$ sideband, are filtered out from the reflection port of FBG with high suppression ratio attributed to its rectangular characteristic. After circulation, that two components with 6.834 GHz frequency gap are optically amplified to several Watts by an erbium-doped fiber amplifier (EDFA), and then pumped into the periodically-poled lithium niobate (PPLN) waveguide mixer for nonlinear frequency conversion. As a result, the 780-nm Raman light determined for atom interferometry is generated by second harmonic generation (SHG) in PPLN with frequency conversion from the filtered carrier and +1$^{st}$ sideband.

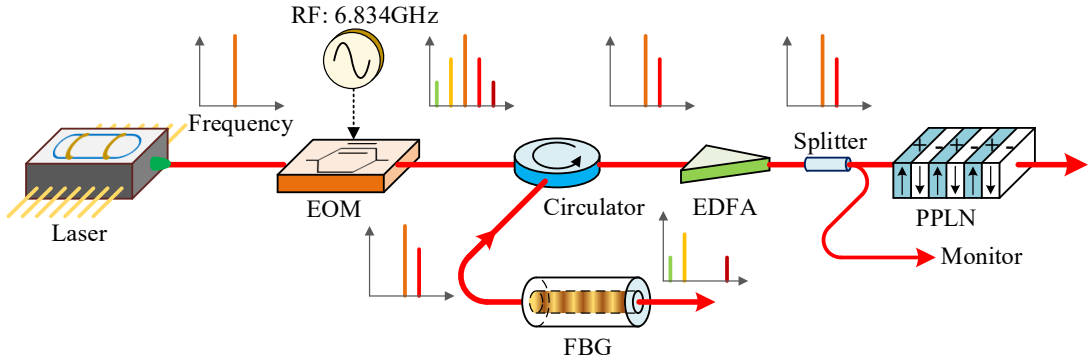

**Fig. 1.** Optical configuration diagram of Raman light generation

### 3.2 Filtering effect of a rectangular FBG

Figure 2 shows an implementation scheme using the designed FBG for the generation of single-sideband-modulated Raman light. Optical frequency of the seed laser is stabilized at $f_1/2$, while $f_1$ corresponds to the low frequency of the Raman light which is 2.4 GHz away from atomic transition $^{85}$Rb: F=3→F'=4 and leads to 1 GHz Raman detuning. After phase modulation, double sidebands arise on the positive and negative sides of the carrier $f_1/2$. By elaborating design parameters of FBG, such as grating pitch and grating length, the reflective spectrum of FBG is tuned to make only the carrier and +1$^{st}$ sideband located at the flat top of the spectrum, and the residual sidebands are enormously suppressed. Based on the power amplification by EDFA and nonlinear frequency conversion by PPLN, three frequency components of 780-nm lasers are generated, the first component of $f_1$ from frequency doubling of the carrier, the second component of $f_1$+6.834 GHz from frequency sum of the carrier and +1$^{st}$ sideband, and the third component of $f_1$+13.668 GHz from frequency doubling of the +1$^{st}$ sideband. The components of $f_1$ and $f_1$+6.834 GHz form the targeted Raman light while the component of $f_1$+13.668 GHz is treated as a harmful residue. As the power ratio of the carrier to the +1st sideband is usually set between 6 and 10 for a practical and optimized Raman light (the power ratio of Raman1 to Raman2 keeps at 2 ± 0.5), the residue accounts for a small proportion compared with the desired Raman light according to the frequency doubling theory in nonlinear crystal [22]. Although the residue along with Raman 2 can also form another pair of Raman light, due to a large frequency detuning and a relative lower power, the influence is tiny, even being eliminated by parameter optimization [11].

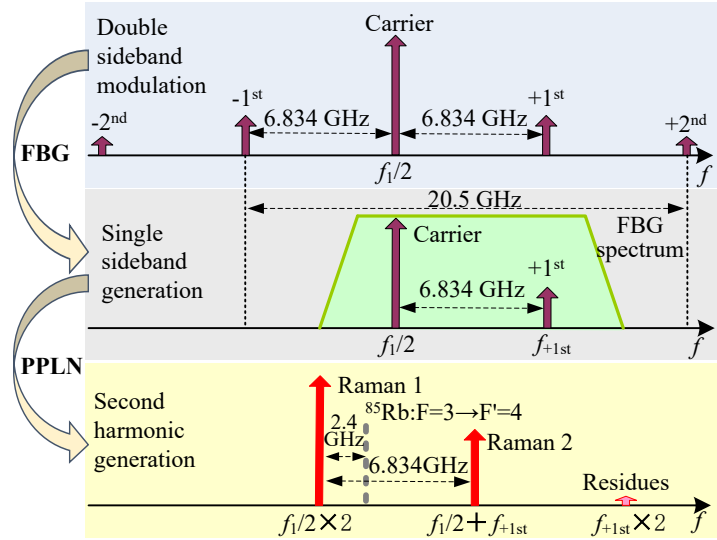

**Fig. 2.** Schematic diagram of single-sideband Raman light generation

## 4. FBG fabrication and performance test

The used FBG is fabricated using ultraviolet exposure post-processing method and inscribing in PM1550 (Corning INC) [23]. As the grating period $\Lambda$ that we used is about 539.2 nm, a center wavelength of 1560.48 nm is produced with fiber refractive index $n$=1.44703, and that it is tunable more than ±150 pm by temperature controlling. In order to achieve the rectangular filtering response and high S/N ratio modulation-sideband filtering effect, the grating length is about 60 mm to achieve a 1dB-bandwidth of about 75 pm, as the FBG bandwidth is dependent on the fiber length. In order to achieve a good rectangular filtering response, the grating was apodized by the **sinc** square function with its main lobe coincided with the grating length, achieving a 23 dB bandwidth less than 160 pm (corresponding to 20 GHz) [24]. An Amplified spontaneous emission (ASE) light source is used as the testing optical source to be input into the fabricated FBG by a fiber circulator, and an optical spectrum analyzer (APEX, AP268XB) with resolution of 0.04 pm is used to measure the reflection spectrum.

The reflective spectrums of FBG are shown in Fig. 3-a, where spectrums under different temperatures are collected in a 3-D diagram with three axes of wavelength, temperature and relative power. The spectrum results show that the 0.8 dB bandwidth is 75 pm, and 23 dB bandwidth is 145 pm, revealing a good rectangular filtering shape. The center wavelength variations are specifically given by spectrum figuration in Fig.

3-c, where changes of the center wavelength are less than 6 pm under the environmental temperature change of more than 30 °C. To confirm the FBG performance for OSSB modulation filtering, we used a single-wavelength fiber laser of 1560.442 nm as the input of the phase-modulated EOM driven by a RF source of 6.834 GHz, and tuned the modulation depth from zero to make the relative power of the carrier and the ±1st sidebands as 1:1:1. After filtering by FBG, we used the AP268XB to acquire the reflected spectrum as shown in Fig. 3-b. From the view of normalized relative power, the carrier power is equal to the +1st sideband power while the -1st sideband power is suppressed by more than 25 dB. The suppression ratio versus the environmental temperature is given in Fig. 3-d, where the suppression ratio fluctuates from -25.8 dB to -30.2 dB under the same temperature change range as above. These results prove that the designed rectangular FBG is very robust and effective to extract the targeted carrier and sideband from double sidebands, succeeding in a remarkable OSSB modulation.

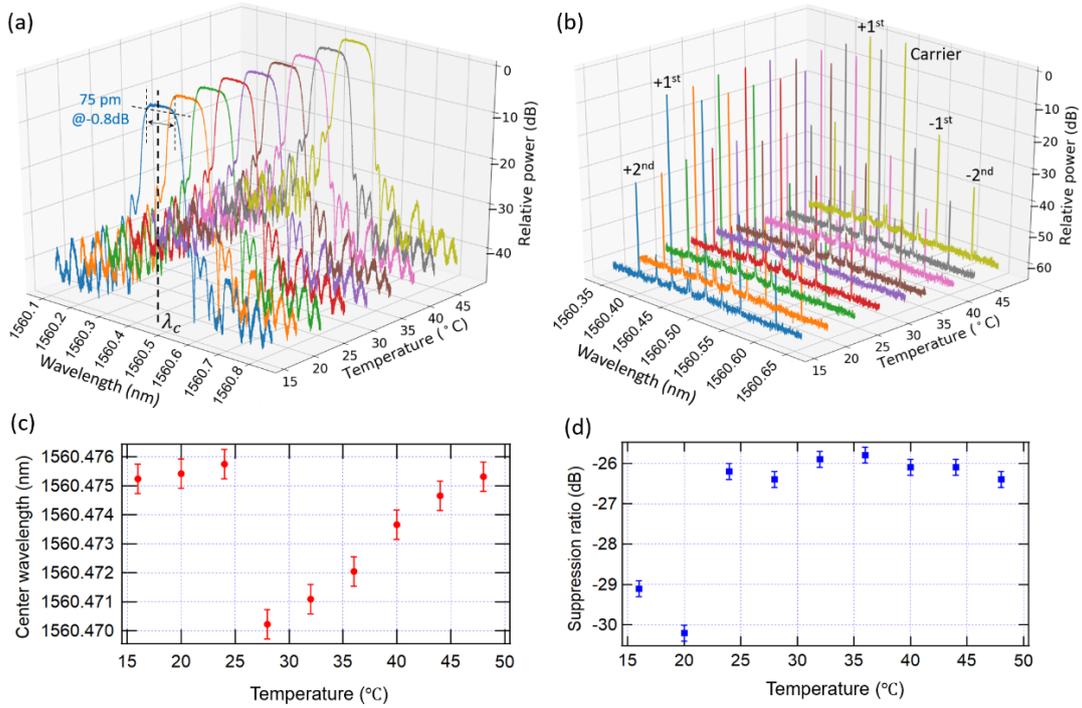

**Fig. 3.** Tested spectrum for FBG. (a) Reflective spectrum of fabricated FBG by ASE. (b) Reflective spectrum by phase-modulated lasers. (c) Center wavelength versus the environmental temperature. (d) Suppression ratio versus the environmental temperature.

## 5. Experimental results
### 5.1 Results of single-sideband Raman light generation

As it has been explained above that the 1560-nm FBG-filtered light is frequency-doubled to 780-nm single-sideband Raman light, but the power ratio between Raman1 and Raman2 is tunable, and the residue intensity will probably rise up. To cancel the first order AC Stark shift, the power ratio ought to be chosen by adjusting the modulation depth of EOM based on physical experiments. After phase-shift location of atom interferometry, two scanning Fabry-Perot interferometers (FPI), a 1560-nm FPI (Thorlabs SA210) with a free spectral range (FSR) of 10 GHz and a 780-nm FPI (Thorlabs SA200) with an FSR of 1.5 GHz, are used to to probe the proper power ratio before frequency doubling and after frequency doubling, respectively. Moreover, to simultaneously observe the power ratios, the two FPIs share the same PZT scanning voltage driver. The spectra of FBG filtered light and the single-sideband Raman light, acquired by Tektronix MSO64 at a sampling rate 1 M/s and with a power ratio of -3 dB between Raman1 and Raman2 for phase-shift compensation, are shown in Fig. 4. Figure 4-a shows the triangular wave scanning voltage for PZT in a period of 0.2 s. Figure 4-b shows the optical spectrum of the laser extracted by FBG, where two clean frequency components are obviously observed and the power ratio of the carrier to the +1$^{st}$ sideband is -8 dB. Figure 4-c shows the optical spectrum of the generated Raman light. Beside the two Raman frequencies spaced 835 MHz apart due to FSR of FPI, the main residual frequency is also observed close to Raman1 resulting from second harmonic generation of the +1$^{st}$ sideband, about 168 MHz apart from Raman1, and this undesired frequency has a relative SNR of lower than -19 dB compared to the power of Raman1.

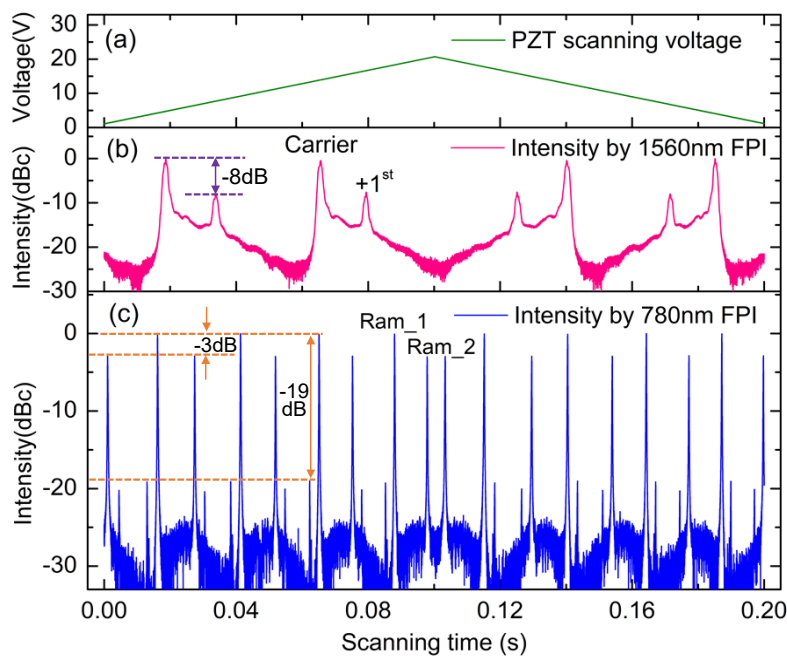

**Fig. 4.** Tested spectrum for single-sideband Raman light. (a) PZT scanning voltages for FPI. (b) FBG filtered spectrum. (c) Spectrum of the single-sideband Raman light.

In order to test the anti-disturbance ability of single-sideband Raman light generation, we deliberately made environmental perturbation onto the EOM-FBG integrated module. This module attached with a temperature sensor probe (Thorlabs-TSP01) was enclosed in a sealing cover, and hot air flow from a hair-dryer was blew into the sealing cover to heat up the whole module. Figure 5-a and 5-b give the 40 s-duration results of the SNR on Raman light generation, corresponding to the relative power ratio of the residual sideband to Raman1, and temperature, respectively. It is notable that the SNR concentrates in the area from -19.5 dB to -18.6 dB, with an average of -19.1 dB and standard deviation of 0.16 dB, while the temperature increases from 27.8 °C to 34.5 °C. This result testifies the robustness of single-sideband Raman light against environmental disturbances.

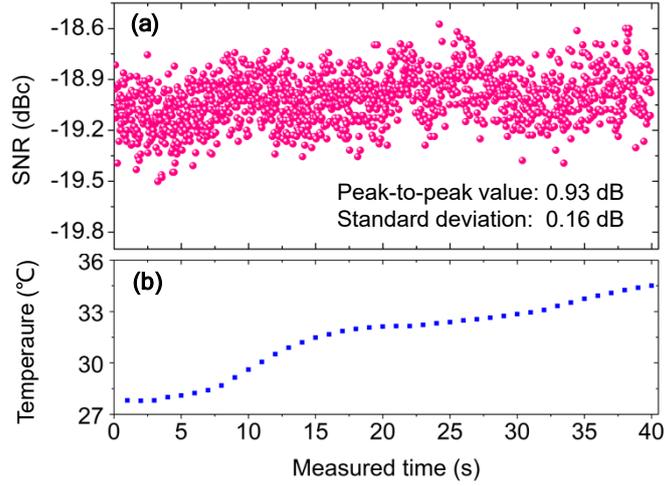

**Fig. 5.** Thermal shock test for FBG-based Raman light generation. (a) SNR of the residual sideband compared to the power of Raman1. (b) Monitored temperature variation for active heating up.

### 5.2 Performance of single-sideband Raman light

The measurement sensitivity of cold atom gravimeters can be degraded by the relative phase noise between the Raman lasers [25]. However, as the optical frequency pair of the Raman light is inherently generated by a phase modulator and shares the same optical path, phase noise of the beat note detected from the Raman light almost perfectly coincides with that of the RF source up to a 1 MHz offset frequency, as is shown in Fig. 6 and measured by a signal analyzer (Anapico APPH20G). The small phase noise bump between 1 kHz and 3 kHz is probably attributed to the PZT bandwidth characteristic of the fiber laser. At offset frequencies of above 500 kHz, the measured phase noise is higher than that of the RF modulation signal due to the intensity noise of the fiber laser. The limitation to the gravity measurement sensitivity caused by the phase noise is on the level of μGal/shot for the Raman-pulse-type atom interferometry [26][27].

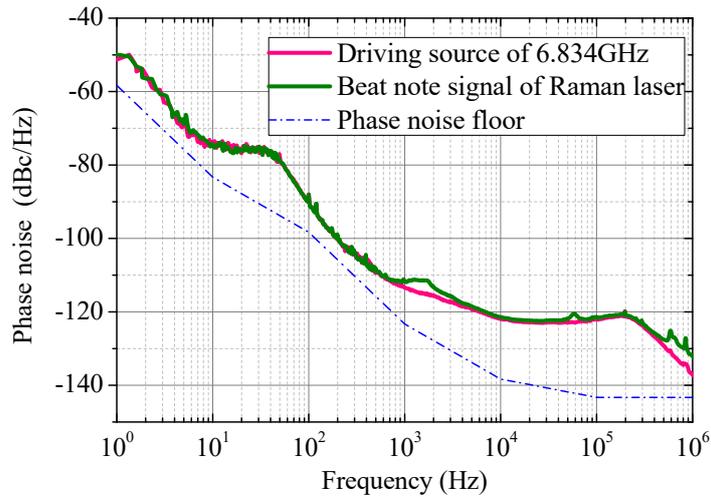

**Fig. 6.** Phase noise analysis of the generated Raman light by Anapico APPH20G

The generated Raman light was then used to perform gravity measurements in a Raman-pulse-type atom interferometer. The atomic cloud was cooled to an initial temperature of about 4μk by magnetic-optical-trap (MOT) technique and polarization gradient cooling (PGC) process, and fell freely under gravity with a vertical length of 21.6 cm to realize the matter-wave interference by three Raman pulse. Finally, the relative population between two ground states was detected to obtain the interference fringe whose phase is closed related to the local gravity. In order to compensate the Doppler frequency shift induced by the free falling of the atoms, the driven frequency for Raman laser was swept at a chirp rate $\alpha$. At a specific chirp rate, the phase shift induced by the gravitational acceleration is canceled by sweeping the frequency gap between the Raman lasers and there exists a stationary phase point independent of interferometric time $T$. Therefore, the value of $g$ can be derived from the frequency chirp rate by the formula of $g = 2\pi\alpha/k_{eff}$, where $k_{eff}$ is the effective wave vector of Raman laser [1]. The average power of the generated Raman light in this work was ~200 mW, and the corresponding Raman π-pulses length was ~13 μs with the single-photon detuning Δ=1GHz relative to the excited state of $|5^2P_{3/2}, F'=1\rangle$. Figure 7 shows the measured interference fringes with $T$ equal to 11ms, 23 ms and 50 ms respectively. The fringe contrast is only about 30% which is limited by the imperfect optimization on parameters, including the residual magnetic fields, the finite size of the cloud and the beam alignment, etc. The local gravity $g$ is initially determined as 9.790095±0.000001 m/s² through three interference fringes for a Doppler compensation. Moreover, the residual sidebands of the generated Raman light in our system probably induce a sub-μGal error according to numerical analysis [28]. This demonstration shows the successful atomic interferences by the generated Raman light with our method, verifying its applicability to atom interferometry.

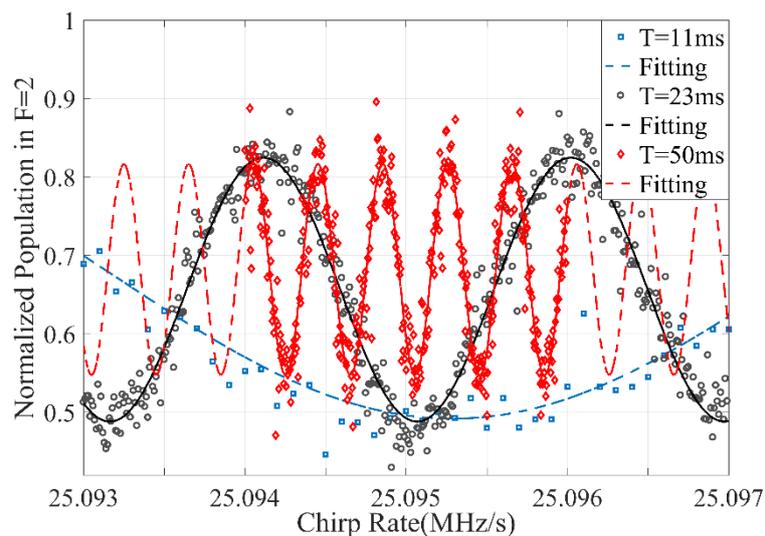

**Fig. 7.** The interferometric fringes for determining local gravity. The chirp rate at the stationary phase point is 25, 095, 055 ±3Hz, from which the local gravity can be derived as 9.790095±0.000001 m/s².

# 6. Conclusion

In conclusion, a single-sideband Raman light scheme based on FBG technology has been successfully demonstrated through the realization of a robust, all-fiber laser system. Remarkable suppression ratios both at 1560 nm and 780 nm have been achieved, namely -26 dB and -19 dB, respectively. On a basis of the Raman light generation, the interferometric fringes and gravity measurement have been demonstrated with Rubidium cold atom ensemble. Compared with other modulated Raman light systems for rubidium atom interferometry, to the best of our knowledge, our scheme for single-sideband-modulated Raman light is most effective and robust in the face of temperature variations. This method and system for Raman light generation can make full use of laser power, be immune to influences on residual sidebands, and be easily fiber-integrated, which facilitates portable and field applications of atom interferometry.

Acknowledgment    This work was supported by National Natural Science Foundation of China (Grant No. 12004428, 62175246), Natural Science Foundation for outstanding young of Hunan Provincial, China (Grant No. 2021JJ20047), China Postdoctoral Science Foundation (Grant No. 2020M683729), and Science Foundation of Hunan Provincial, China (Grant No. 2021JJ30774), Natural science foundation of Shanghai (22ZR1471100).

Disclosure    The authors declare no conflicts of interest related to this paper.